\documentclass[final,3p,times]{elsarticle}

\usepackage{graphicx}
\usepackage{subfigure}

\usepackage{amssymb}
\usepackage{amsmath,bm}
\usepackage{color}
\usepackage{xcolor,colortbl}

\usepackage{lineno}

\journal{Physica A: Statistical Mechanics and its Applications}

\begin{document}
\begin{frontmatter}

\title{On multifractals: a non-linear study of actigraphy data}

\author[label1,label2]{Lucas Gabriel Souza Fran\c{c}a\corref{cor1}}
\cortext[cor1]{Corresponding author}
\ead{lucas.franca.14@ucl.ac.uk}
\author[label3]{Pedro Montoya}
\author[label1,label4]{Jos\'e Garcia Vivas Miranda\corref{cor2}}
\cortext[cor2]{Corresponding author}
\ead{vivas@ufba.br}
\address[label1]{Department of Physics of the Earth and the Environment, Institute of Physics, Federal University of Bahia, Salvador, Brazil}
\address[label2]{Department of Clinical and Experimental Epilepsy, UCL Queen Square Institute of Neurology, University College London, London, UK}
\address[label3]{Research Institute of Health Sciences (IUNICS), University of the Balearic Islands, Palma de Mallorca, Spain}
\address[label4]{Motion Analysis Laboratory, Spaulding Rehabilitation Hospital, Harvard Medical School, Boston, MA, USA}

\begin{abstract}
The healthcare process generates a vast quantity of data. Exploitation and analysis of such material might lead discoveries and potentially improve the clinical activity. Over the last years, several studies focused on the non-linear features exhibited by physiological signals and medical data. In general, the complex features of these signals have been demonstrated in some studies, ranging from heartbeat time intervals to neuron spikes series. Some works employed fractal and multifractal concepts to the analysis of such medical data. Similar evaluations of movement patterns recorded by actigraphy devices on individuals with fibromyalgia might provide new information that can be potentially translated into the clinical practice. Chronic pain is a condition characterised by sleep disturbances and psychological disorders that is correlated with impairment and reduced physical activity. This work aimed to determine the characteristics of activity series from fractal geometry concepts, in addition to evaluate the possibility of identifying individuals with fibromyalgia. Activity level data were collected from 27 healthy subjects and 27 fibromyalgia patients, with the use of clock-like devices equipped with accelerometers, for about two weeks, all day long. The activity series were evaluated through fractal and multifractal methods. Hurst exponent analysis exhibited values according to other studies for both groups, however, it is not possible to distinguish between the two groups by such analysis. Activity time series also exhibited a multifractal pattern. A paired analysis of the spectra indices for the sleep and awake states revealed differences between healthy subjects and fibromyalgia patients. The individuals feature differences between awake and sleep states having statistically significant differences for $\alpha_{q-} - \alpha_{0}$ in healthy subjects only, suggesting that there are not differences between awake and sleep state for patients with fibromyalgia. The approach has proven to be an option on the characterisation of such kind of signals and was able to (indirectly) differ between both healthy and fibromyalgia groups. This outcome suggests changes in the physiologic mechanisms of movement control.
\end{abstract}


\end{frontmatter}
%
%

\section{Introduction}
\label{int}
The medical sciences and health care process generate a vast quantity of data. Data analysis in health sciences could reveal new relationships between symptoms and physiological mechanisms \cite{Prather1997}. The natural way to parse this information comprehends starting with linear methodologies, such as mean and standard deviation, nevertheless linear approaches they do not appropriately handle the enormous inter and intra-individual variability of physiological and psychological symptoms.
\\
Over the last years, several studies focused on the non-linear features exhibited by physiological signals, in general. The complex features of such signals have been demonstrated in some studies, ranging from heartbeat time intervals \cite{Peng1995} to neuron spikes series \cite{Natarajan2004}. They are expressed in heteroscedastic recordings, exhibiting huge variability over time and across different individuals.
\\
Such studies provided new insights on how human control mechanisms work and how the disease processes change these features. In an example of application, researchers evaluated long-term correlation on beat intervals and noticed a variation on the scalling properties in patients suffering with heart diseases \cite{Peng1995}. A further study suggested a multifractal behaviour on the human heartbeat patterns \cite{Ivanov1999}. In both cases, changes in physiologic mechanisms of control, revealed by the multifractal analysis, are pointed as the cause of such changes.
\\
The multifractal geometry is an extension of the fractal definition, that allows a better characterisation of elements that require an infinity number of parameters to be described \cite{Stanley1999}, and is a possible tool to expand knowledge about the dynamics of physiological signals and healthcare data. A potentially eminent application of such approach is the study of actigraphy signal patterns in humans. Such analysis could reveal underlying features of movement in activity recordings.
\\
Actigraphy is a technique that allows the inference of wakefulness and sleep from the presence or absence of movement \cite{Lichstein2006,Chialvo2015,McGowan2018}. Data obtained with such technique were used to study the scaling properties on the activity time series for healthy subjects \cite{Hu2004}. The authors have shown that there are fractal properties in human activity, suggesting a multi-scale mechanism of the locomotor regulation. Later, it was revealed that the activity time series for Alzheimer's disease patients also have scaling properties, however, with a different exponent \cite{Hu2009}. This analysis can uncover new properties of the movement mechanisms that were not accessible with the conventional linear methods.
\\
Applying a multifractal approach to study movement patterns in patients with chronic pain, e.g., fibromyalgia \cite{Wolfe1990,Costa2017}, might supply valuable information about the heterogeneity in the motor behaviour of such group, providing new insights that could be potentially translated into clinical practice.\\
The subjective diagnosis protocol and the lack of a more precise approach to identify the syndrome makes the clinical treatment of affected patients more difficult. A better way to classify and identify potential markers of the syndrome would improve healthcare protocols and provide a better quality of life to patients, as well as, give insights about the aetiology of the illness.\\
The fatigue reported by patients with fibromyalgia (FM) is severe enough to change the kinematic limits of movements, modifying their daily activities with direct impact on quality of life \cite{Bennett2007,Jones2008}. Actigraphy has been used to assess changes in these patterns of activity for some patients \cite{Kashikar-zuck2010,Kashikar-Zuck2013,Korszun2002,Okifuji2011,Sadeh2002}. Moreover, it was shown that actigraphy is a useful way of studying activity levels and sleep patterns, and, have demonstrated significant differences between fibromyalgia patients with and without comorbid depression \cite{Korszun2002}. In all these studies the methods used to characterise the activity were always linear and conclusions based on parametric tests of motion measures.\\
To date, actigraphy data scaling features have not been widely studied. Monofractal properties of actigraphy signals from patients with fibromyalgia have not been explored, so far, as well as the multifractal features of such data in both healthy individuals and patients. This study, then, proposes to analyse these questions in detail and applies monofractal and multifractal methods to actigraphy data in patients with fibromyalgia and healthy subjects looking for markers of the chronic condition.\\
The objective of this work is, then, to evaluate actigraphy data based on multifractal methods, in order to look for motor markers of the syndrome. In this study, the Chhabra and Jensen approach \cite{Chhabra1989} was used to study the scaling properties in activity time series of healthy (HC) and patients with fibromyalgia (FM). In our hypothesis, FM patients should exhibit different scaling properties when compared to healthy individuals.\\

\section{Methodology}
\label{meth}

\begin{figure}[ht]
    \centering
    \includegraphics[width=\columnwidth]{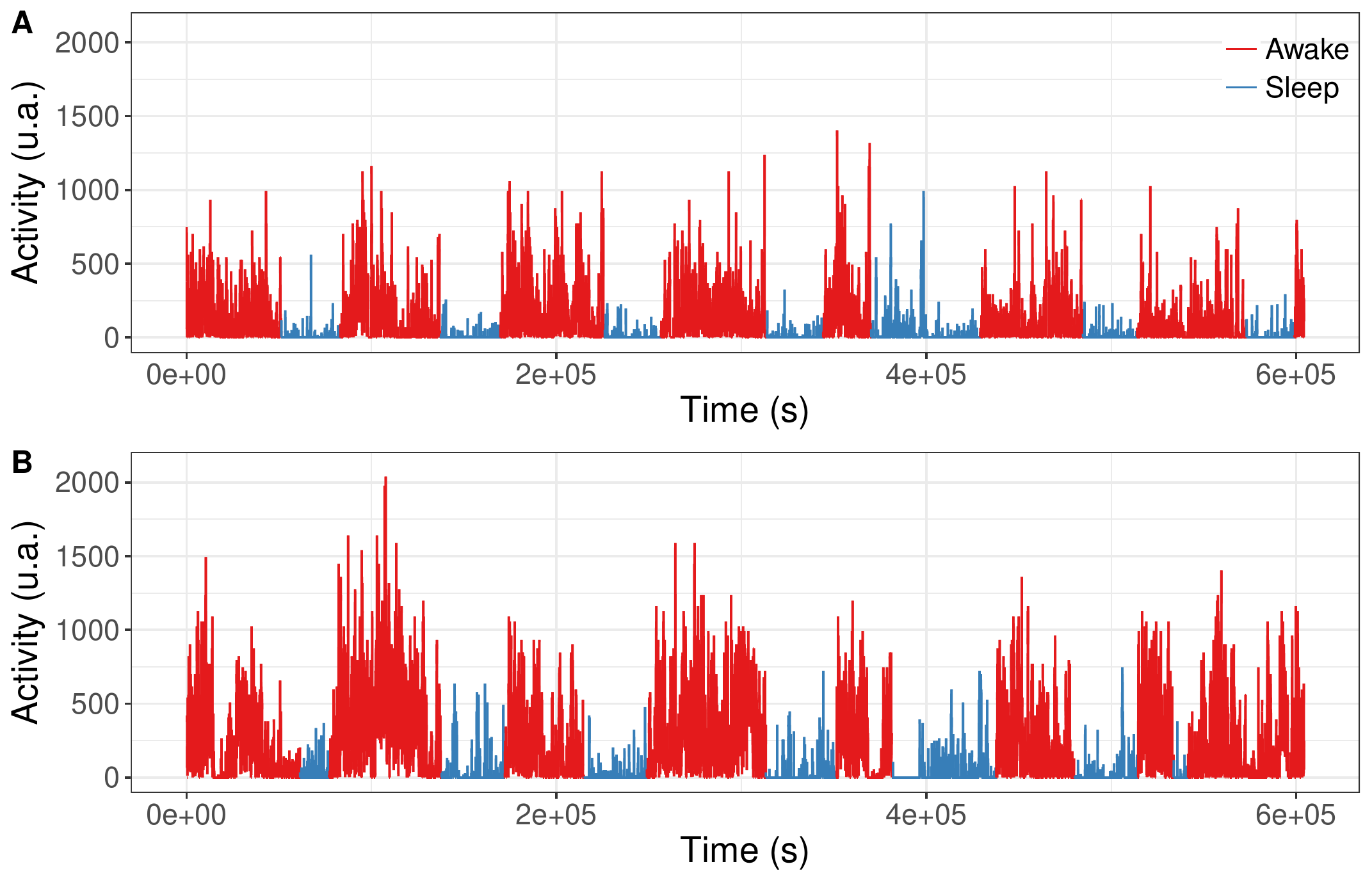}
    \caption{Actigraphy time series showing both states for a healthy control (A) and an individual with fibromyalgia (B). The recordings show fluctuactions on the movement level in unities of acceleration (u.a.). The different states present distinct visual patterns for both subjects, being the amplitude the most notable change.}
    \label{fig:states}
\end{figure}

An actigraphy device, attached to wrist on fifty-four individuals, collected data of physical activity. Individuals regularly wore the devices during a period of about two weeks without interruption, except for bath time. Written inform consent was obtained from all participants and the protocol of the study was approved by the official ethical committee of the Government of Balearic Islands (reference: 1431/10 PI). Individuals were classified in two different groups: patients suffering from chronic pain for more than one year (27 individuals) and healthy subjects (27 individuals). More information about the patients can be found in table \ref{patData}.
The series of activity were split into smaller segments consisting of two possible states for an individual: awake and sleep. The information about the state is given by the actigraphy device and determined by a commercial algorithm. The device registers the activity level at every 30 seconds. After the segmentation, the parts of the same state were stuck together, and then, two series, corresponding to both states, awake and sleep, were created for each individual. Examples of actigraphy for a healthy individual and a patient are shown in figure \ref{fig:states}.

\begin{table}[]
\centering
\caption{Patient information. The painkillers, anxiolytics, and antidepressants columns state if drugs of each category were taken by each individual. F: Female. HC: Healthy Controls. FM: Fibromyalgia Patients. BMI: Body Mass Index. NA: Not Available}
\begin{tabular}{| c | c | c | c | c | c | c | c | c |}
\hline
Patient & Gender & Age   & BMI   & Group & Painkillers & Anxiolytics & Antidepressants \\
\hline
000     & F      & 28.76 & 23.53 & HC    & No           & No           & No               \\
030     & F      & 35.86 & 22.50 & HC    & No           & No           & No               \\
040     & F      & 24.68 & 21.63 & HC    & No           & No           & No               \\
050     & F      & 25.01 & 18.96 & HC    & No           & No           & Yes               \\
060     & F      & 30.29 & 20.94 & HC    & No           & No           & No               \\
070     & F      & 36.02 & 20.76 & HC    & No           & No           & No               \\
080     & F      & 26.33 & 20.69 & HC    & No           & No           & No               \\
100     & F      & 44.70 & 24.84 & HC    & No           & No           & No               \\
110     & F      & 41.88 & 21.49 & HC    & No           & No           & No               \\
120     & F      & 31.94 & 18.71 & HC    & No           & No           & No               \\
140     & F      & 48.69 & 25.56 & HC    & No           & No           & No               \\
150     & F      & 42.61 & 18.86 & HC    & No           & No           & No               \\
160     & F      & 46.66 & 24.51 & HC    & No           & No           & No               \\
001     & F      & 56.97 & 25.59 & HC    & No           & No           & No               \\
011     & F      & 57.69 & 23.59 & HC    & No           & No           & Yes               \\
021     & F      & 54.64 & 25.63 & HC    & No           & No           & No               \\
031     & F      & 56.14 & 22.31 & HC    & No           & No           & No               \\
041     & F      & 71.75 & 24.34 & HC    & No           & No           & No               \\
051     & F      & 57.25 & 26.78 & HC    & No           & No           & No               \\
061     & F      & 58.78 & 25.63 & HC    & No           & No           & No               \\
071     & F      & 57.34 & NA    & HC    & No           & No           & No               \\
091     & F      & 70.11 & NA    & HC    & No           & No           & No               \\
101     & F      & 61.36 & 21.23 & HC    & No           & No           & No               \\
111     & F      & 57.26 & 20.40 & HC    & No           & Yes           & Yes              \\
121     & F      & 58.97 & 20.20 & HC    & No           & No           & No               \\
141     & F      & 57.26 & 21.22 & HC    & No           & No           & No               \\
151     & F      & 57.40 & 23.71 & HC    & No           & No           & No               \\
500     & F      & 43.17 & 29.02 & FM    & Yes          & Yes          & Yes              \\
510     & F      & 41.81 & 30.48 & FM    & Yes          & Yes          & Yes              \\
520     & F      & 40.98 & 27.92 & FM    & Yes          & No           & Yes               \\
530     & F      & 31.13 & 29.03 & FM    & Yes          & Yes          & Yes              \\
560     & F      & 41.71 & 20.58 & FM    & Yes          & Yes          & No               \\
570     & F      & 34.51 & 23.71 & FM    & Yes          & No           & No               \\
580     & F      & 29.97 & 23.34 & FM    & Yes          & No           & No               \\
600     & F      & 38.52 & NA    & FM    & NA          & NA          & NA              \\
610     & F      & 41.99 & 24.45 & FM    & Yes          & Yes          & Yes              \\
620     & F      & 37.56 & 35.42 & FM    & Yes          & Yes          & Yes              \\
630     & F      & 46.57 & 20.90 & FM    & Yes          & Yes          & Yes              \\
640     & F      & 49.04 & 23.62 & FM    & No           & No           & No               \\
650     & F      & 46.12 & 27.04 & FM    & Yes          & Yes          & Yes              \\
501     & F      & 60.71 & 20.03 & FM    & Yes          & Yes          & Yes              \\
511     & F      & 61.83 & 22.84 & FM    & Yes          & Yes          & Yes              \\
521     & F      & 50.47 & 24.97 & FM    & NA          & NA          & NA              \\
531     & F      & 58.99 & 25.54 & FM    & Yes          & Yes          & Yes              \\
541     & F      & 56.39 & 25.89 & FM    & Yes          & Yes          & Yes              \\
\hline
\end{tabular}
\label{patData}
\end{table}

\begin{table}[]
\centering
\begin{tabular}{| c | c | c | c | c | c | c | c | c |}
\hline
Patient & Gender & Age   & BMI   & Group & Painkillers & Anxiolytics & Antidepressants \\
\hline
551     & F      & 52.24 & 25.53 & FM    & No           & Yes          & Yes              \\
561     & F      & 46.62 & 27.64 & FM    & Yes          & Yes          & Yes              \\
571     & F      & 58.31 & 34.93 & FM    & Yes          & No           & Yes              \\
581     & F      & 50.50 & 36.15 & FM    & Yes          & Yes          & Yes              \\
601     & F      & 56.45 & 28.00 & FM    & Yes          & Yes          & Yes              \\
611     & F      & 60.93 & 23.05 & FM    & Yes          & Yes          & Yes              \\
621     & F      & 54.94 & 27.82 & FM    & Yes          & Yes          & Yes              \\
631     & F      & 57.63 & 39.04 & FM    & Yes          & Yes          & Yes              \\
641     & F      & 56.95 & 26.04 & FM    & Yes          & No           & No              \\
\hline
\end{tabular}
\end{table}
Fractal analysis was performed on original series and multifractal analysis on the segmented states series. The Hurst exponent was estimated with the Detrended Fluctuation Analysis (DFA) method \cite{Peng1994}, and an algorithm available on the PhysioNet repository \cite{Goldberger2000}. Parameters for the fractal analysis followed the previous ones found in literature \cite{Hu2004}, with a window size varying from 2 to 250 minutes.
For the multifractal analysis, the method developed by Chhabra and Jensen \cite{Chhabra1989} was used due to its accuracy and implementation easiness. The method enables a precise determination of the generalized dimensions $D_{q}$ and $f(\alpha)$ singularity spectra, avoiding disadvantages of using a Legendre transform, employed in some methods like the Multifractal Detrended Moving Average (MFDMA) approach \cite{Gu2010}, and it is suitable for experimental data of a system of an unknown dynamics \cite{Chhabra1989}.

The generalized dimensions, $D_{q}$, evaluate the $q_{th}$ order scaling properties of the profile and is defined in equation \ref{genDSpectrum}.

\begin{equation}
\begin{gathered}
    D_{q} = \frac{1}{q-1}\lim_{\epsilon\rightarrow0}{\frac{log{\sum}_{i}{P_{i}^q(\epsilon)}}{\log{\epsilon}}}
    \label{genDSpectrum}
\end{gathered}
\end{equation}

For $q = 1$, $D_{1}$ becomes indeterminate and can be obtained via eq. \ref{genDSpectrumq1} \cite{Murcio2015}.

\begin{equation}
\begin{gathered}
    D_{1} = \lim_{\epsilon\rightarrow0}{\frac{\sum_{i}{P_{i}(\epsilon)\log{P_{i}}(\epsilon)}}{\log{\epsilon}}}
    \label{genDSpectrumq1}
\end{gathered}
\end{equation}

Where $P_{i}(\epsilon)$ represents the probability of the measure in the $i^{th}$ segment of size $\epsilon$, calculated by dividing the sum of the activity value in the segment $i$ by the sum in the whole support length. $q$ acts as a filter for large measures $(q>0)$ and for small measures $(q<0)$. For $q=0$, $D_{0}$ indicates the capacity dimension, representing the scaling of how measures are distributed in the support. In actigraphy data this index characterises the irregularity of how individuals move or not independently of the intensity.\\
The $f(\alpha)$ singularity spectra are defined parametrically in q as $f(q)$ and $\alpha(q)$. In order to estimate the multifractal spectra, the $qth$ order normalised probability measures $\mu_{i}(q,\epsilon)$ are derived according to the equation \ref{partFunc}. 

\begin{equation}
    \mu_{i}(q,\epsilon) = \frac{[P_{i}(\epsilon)]^q}{\sum_{i}{[P_{i}(\epsilon)]^q}}
    \label{partFunc}
\end{equation}

The Chhabra and Jensen \cite{Chhabra1989} definitions can be expressed as shown in the equations \ref{alphaQ} and \ref{fQ}.

\begin{equation}
    \alpha(q) = \lim_{\epsilon\rightarrow0}\frac{\sum_{i}{\mu_{i}(q,\epsilon)\log{P_{i}(\epsilon)}}}{\log{\epsilon}}
    \label{alphaQ}
\end{equation}

\begin{equation}
    f(q) = \lim_{\epsilon\rightarrow0}{\frac{\sum_{i}{\mu_{i}(q,\epsilon)\log{\mu_{i}(q,\epsilon)}}}{\log{\epsilon}}}
    \label{fQ}
\end{equation}

\begin{figure}[ht]
    \centering
    \includegraphics[width=\columnwidth]{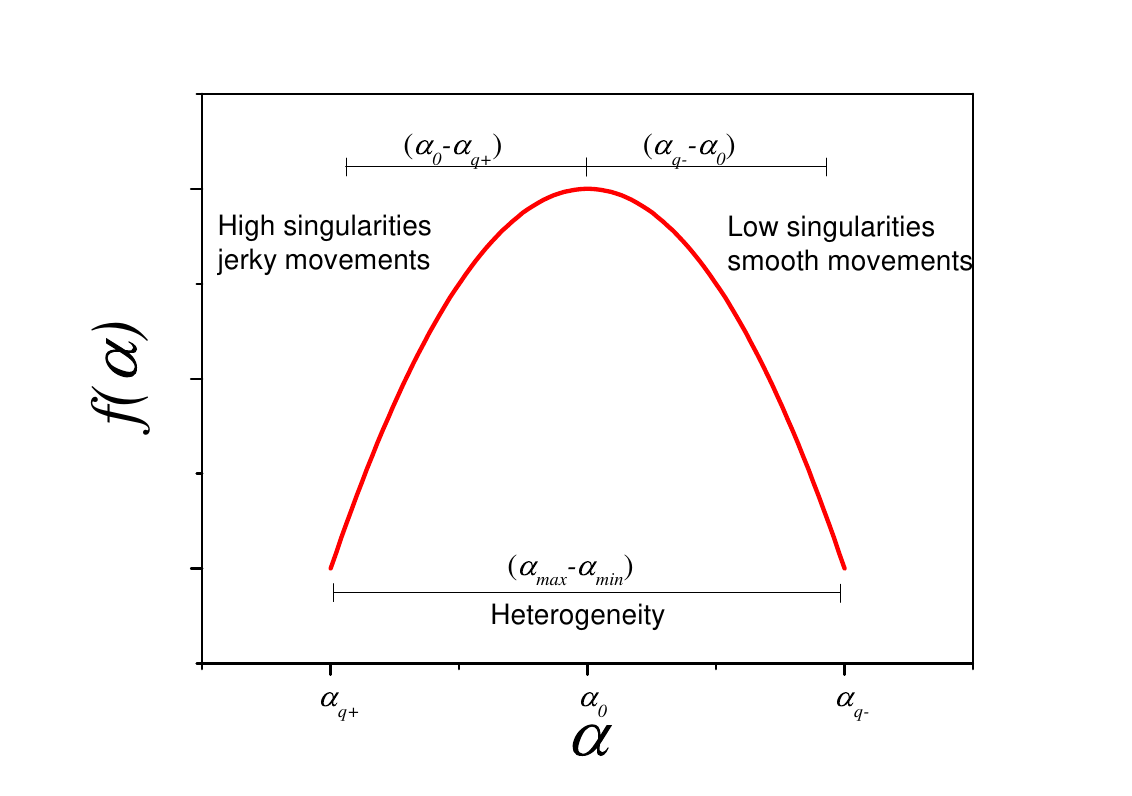}
    \caption{Multifractal spectrum and meaning of the multifractal indices. $(\alpha_{max} - \alpha_{min})$ measures the heterogeneity in the time series, as $(\alpha_{q+} -  \alpha_{0})$ and $(\alpha_{0} -  \alpha_{q-})$ assess such quantity for high and low singularities, respectively.}
    \label{fig:multfr_idx}
\end{figure}

Multifractal indices are powerful tools for characterising scaling properties of time series. The difference $(\alpha_{max} - \alpha_{min})$ is used to examine the heterogeneity in the scaling properties. For monofractals profiles, this difference is approximately zero. In actigraphy measures, this difference indicates the heterogeneity of how smooth or jerky movements are.\\
Low values of $\alpha$ are related with jerking movements and high values with smooth ones. $f(\alpha)$ estimate the scaling exponent for each singularity $\alpha$. This heterogeneity index can be divided in two indices: for low values of $\alpha$, $(\alpha_{q+} -  \alpha_{0})$ and for high values $(\alpha_{0} -  \alpha_{q-})$, where $\alpha_{q-}$ and $\alpha_{q+}$ are the $\alpha$ values for the minimum and maximum value of $q$ considered, respectively. A diagram illustrating the indices meaning is presented in figure \ref{fig:multfr_idx}.

For this method, in order to estimate the partition function, a range of scales $(\epsilon)$ between 4 and 512 points was chosen. It corresponds to a time interval between 2 and 256 minutes. The $q$ value varied from -25 to 25 in steps of 0.3. Values of $q$ featuring fittings with $R^{2} < 0.7$ were rejected. The multifractal indices ($\alpha_{max} - \alpha_{min}$), ($\alpha_{0} - \alpha_{q+}$), ($\alpha_{q-} - \alpha_{0}$), $D_{0}$, $D_{1}$ and $D_{2}$, were estimated for each activity state (awake and sleep) for every subject. A paired t-test was then performed to verify statistical differences between awake and sleep states in both HC and FM.

The original scripts used in this work are available in the public software repository GitHub \cite{actiMF}. In addition, the following software packages were used: R \cite{R2018}, reshape \cite{Wickham2007}, stringr \cite{Wickham2018}, RColorBrewer \cite{Neuwirth2014}, cowplot \cite{Wilke2017}, ggplot2 \cite{Wickham2009}, and boot \cite{Davison1997,Canty2017}.

\section{Results}
\subsection{Monofractal analysis}
\begin{figure}[ht]
    \centering
    \includegraphics[width=\columnwidth]{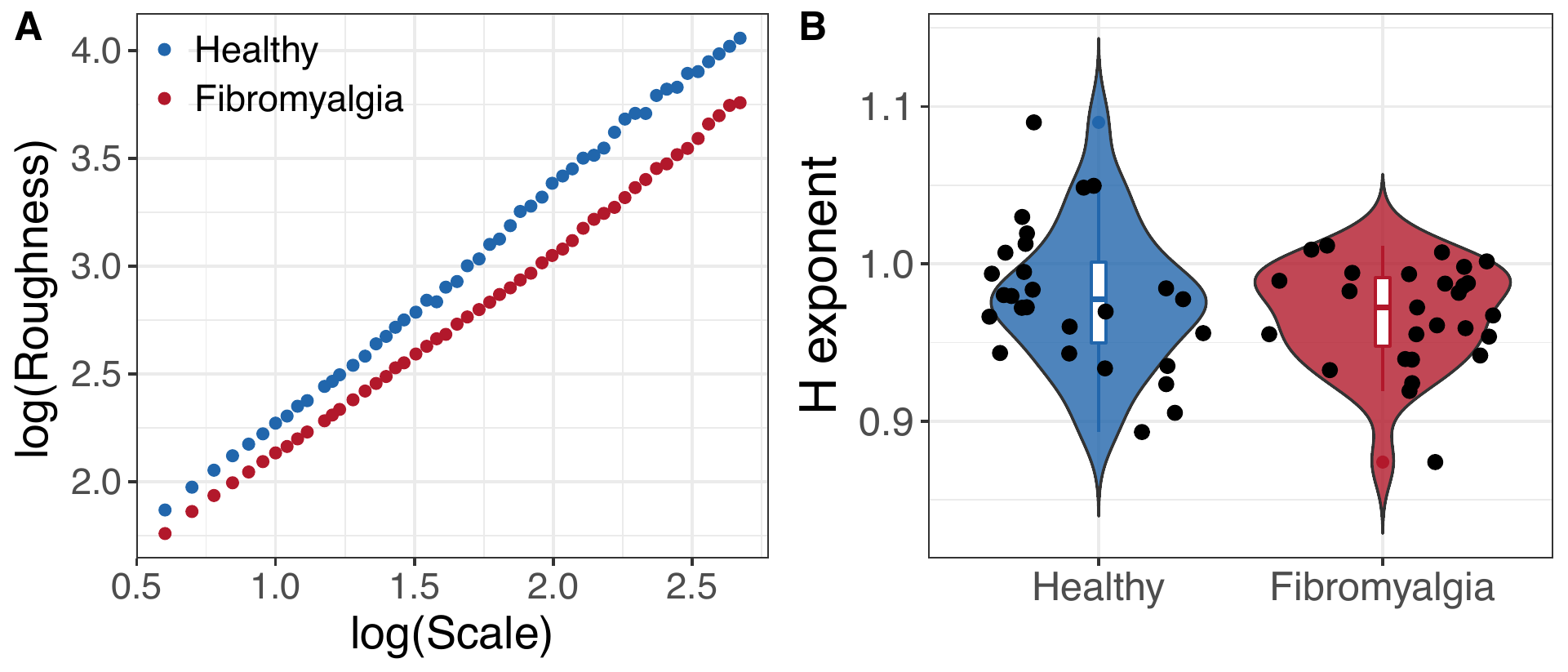}
    \caption{Monofractal analysis of actigraphy data for both healthy and fibromyalgia groups. (A) Roughness versus scale for two individuals (one of each group). (B) The distributions are similar, i.e., they have approximate mean, and present values on the persistent behaviour exponents range.}
    \label{fig:hurst}
\end{figure}

The monofractal analysis revealed a persistent behaviour ($H>0.5$) for both groups ($H=0.98\pm0.04$ for healthy subjects and $H=0.97\pm0.03$ for fibromyalgia patients). However, it is not possible to differ between fibromyalgia patients and healthy subjects by a monofractal analysis, as it can be seen in figure \ref{fig:hurst}. A t-test was performed to compare the monofractal indices for both groups and has not shown significant differences in the results with $p = 0.303$ and $T = -1.04$.

\subsection{Multifractal analysis}
\begin{figure}[ht]
\includegraphics[width=\columnwidth]{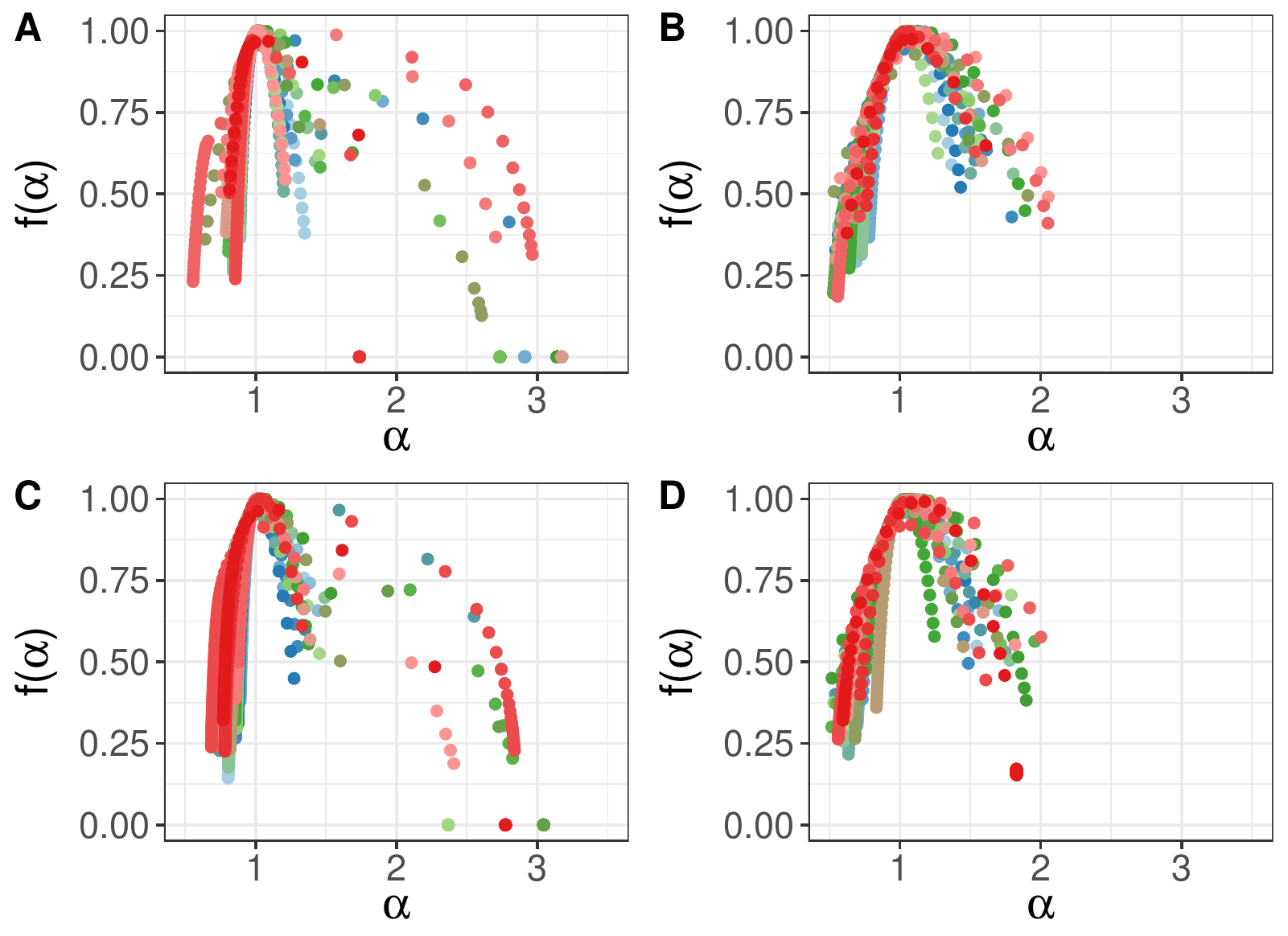}
\caption{Multifractal spectra derived for both groups (HC (A and B) and FM (C and D) and states (awake (A and C) and sleep (B and C)). The main feature is the variability of the curves, i.e, individuals present different spectral shapes regardless of their groups. The curves for sleep state show more symmetrical shapes.}
\label{fig:spectra}
\end{figure}

The multifractal spectra derived with the Chhabra-Jensen method revealed wide and distinct shapes for the different states. Moreover, such curves presented variability between the individuals, i.e., visually they do not seem to be defined by the group - HC or FM - but by each individual. The spectra obtained for every person of both groups can be seen in figure \ref{fig:spectra}.

\begin{figure}[ht]
    \centering
    \includegraphics[width=\columnwidth]{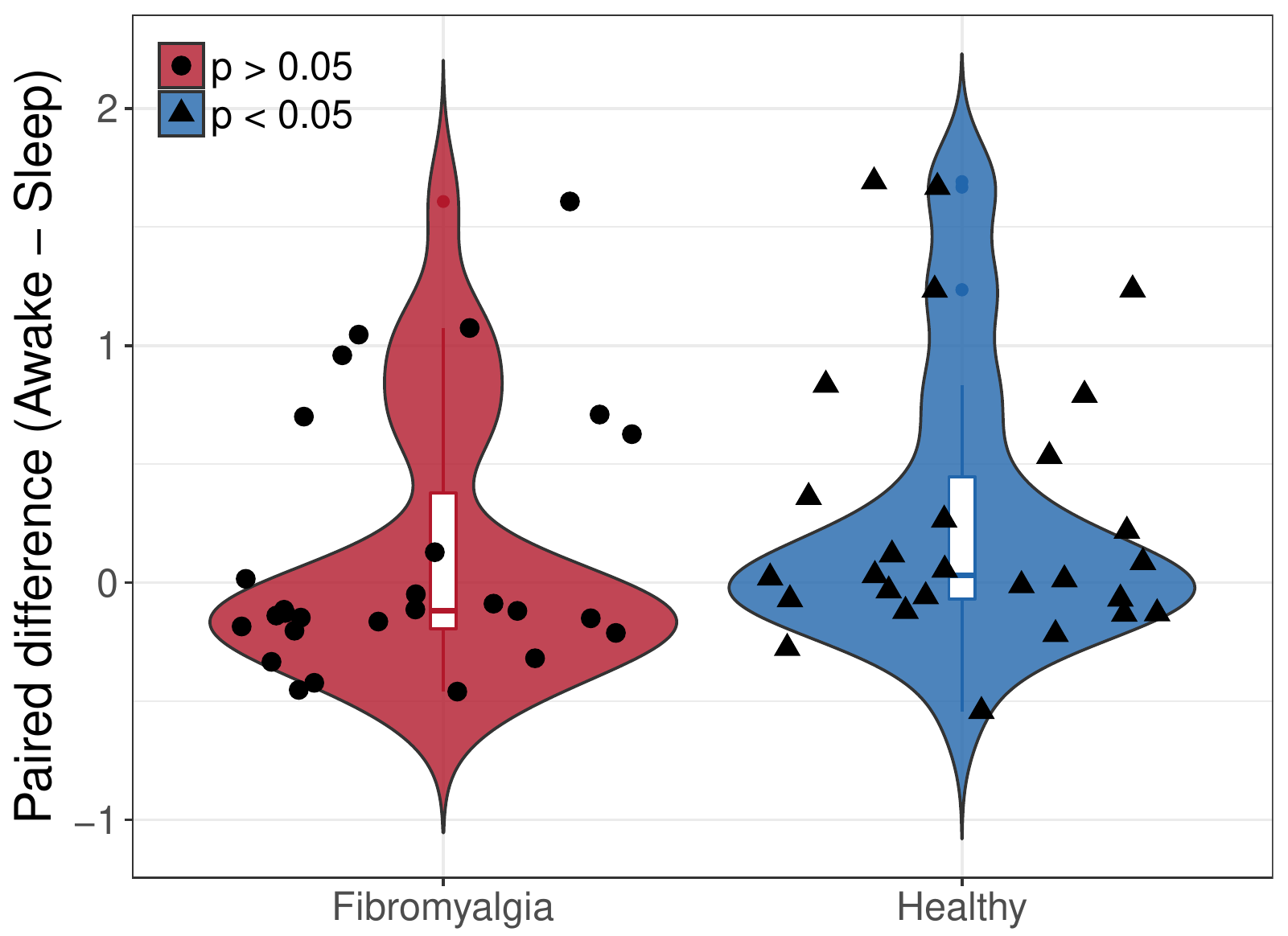}
    \caption{Differences between ($\alpha_{q-} - \alpha_{0}$) for the awake and sleep states. Each point in the graph represents the difference in the index between the awake and sleep states for one individual. A paired t-test shows statistically significant differences between awake and sleep measures for the healthy group (in blue). Healthy individuals present changes on the variable ($\alpha_{q-} - \alpha_{0}$), implying in changes for smooth movements when comparing awake and sleep states.}
    \label{fig:paired_diff}
\end{figure}

No differences were found on comparing the healthy subjects and fibromyalgia patients group, however, when comparing the difference between states of daytime activity and sleep in the same subject, in a paired t-test, $\alpha_{0} - \alpha_{q+}$, $D_{0}$, $D_{1}$, and $D_{2}$ showed statistically significant differences between the awake and sleep states for both HC and FM groups. The spectrum width $\alpha_{q-} - \alpha_{q+}$ has not shown any differences for either groups.
The right side of the parabola, however, only shows statistically significant differences between awake and sleep states for healthy controls with $p = 0.021$. Figure \ref{fig:paired_diff} and table \ref{resumo_testes} show the results for a paired t-test comparing these multifractal indices. For healthy (HC) subjects, exclusively, there are significant differences between daytime and sleep activity for smooth movements ($\alpha_{q-} - \alpha_{0}$). 
A bootstrap performed with a 1000 resamples (with repetion) \cite{Davison1997,Canty2017} resulted in 11\% of p-values $p < 0.05$ for patients with fibromyalgia, whilst healthy individuals returned 70\% of the resamples with $p < 0.05$.

\begin{table}[]
    \centering
\caption{Results of paired t-tests comparing multifractal spectrum measures $\alpha_{0} - \alpha_{q+}$, $\alpha_{q-} - \alpha_{0}$, $\alpha_{q-} - \alpha_{q+}$, $D_{0}$, $D_{1}$, and $D_{2}$ for both awake and sleep states, in healthy subjects and individuals with fibromyalgia. FM and HC show similar results all the quantities, except the spectrum width $\alpha_{q-} - \alpha_{q+}$, and the right side of the parabola ($\alpha_{q-} - \alpha_{0}$) for individuals with fibromyalgia. In this case, only healthy individuals show statistically significant differences. *$p<0.05$, ***$p<0.001$.}
\definecolor{LightCyan}{rgb}{0.88,1,1}

\begin{tabular}{| c | c | c |}
\hline
Index & FM & HC\\
\hline
$\alpha_{0} - \alpha_{q+}$ & $p = 6.925 x 10^{-6}$*** $T = -5.60$ & $p = 0.004x10^{-1}$*** $T = -4.08$\\
\rowcolor{LightCyan}
$\alpha_{q-} - \alpha_{0}$ & $p = 0.293$ $T = 1.07$ & $p = 0.021$* $T = 2.45$\\
$\alpha_{q-} - \alpha_{q+}$ & $p = 0.383$ $T = -0.89$& $p = 0.686$ $T = 0.41$\\
$D_{0}$ & $p = 0.003x10^{-1}$*** $T = -4.16$& $p = 0.020$* $T = -2.48$\\
$D_{1}$ & $p = 2.212x10^{-7}$*** $T = 6.95$& $8.639x10^{-7}$*** $T = 6.41$\\
$D_{2}$ & $p = 9.394x10^{-10}$*** $T = 9.30$& $2.337x10^{-9}$*** $T = 8.89$\\
\hline
\end{tabular}
\label{resumo_testes}
\end{table}

\section{Discussion}
\label{disc}

Monofractal analysis results, for both groups, exhibit a value according to literature \cite{Hu2004}, however, it is not possible to differ between the two populations by such analysis. The monofractal evaluation is not sensitive to the changes in the movement activation pattern; the result indicates the need of a new framework, such as multifractal analysis, that has been extensively employed on physiological signals studies \cite{Stanley1999}.\\
Such approach might provide important practical diagnostic and prognostic not obtainable with the canonical methods. The versatility of the technique when it comes to the wide range of different types of signal to which it can be applied is another important feature \cite{Stanley1999}.\\ 
The multifractal approach was applied in different studies on physiologic signals, e.g., in human heartbeat dynamics \cite{Ivanov1999}, evidencing differences between healthy individuals and patients with heart disease; seizure occurrence \cite{Zhang2015}, on a trial to build a detection protocol; and in postural sway study \cite{CliffordSTChoy2015}.\\
In this study, the multifractal spectra derived with Chhabra-Jensen method \cite{Chhabra1989} exhibited a large variability across different individuals, suggesting an individual component of the scaling properties in actigraphy recordings and, consequently, in physiologic mechanisms of control. A paired analysis, comparing both states, was employed to bypass the limitations associated with the individual variability. Such analysis should be able remove the individual features that difficult the characterisation of the phenomenon and exhibit changes caused by the syndrome.
The results present differences between FM patients and healthy individuals, when comparing awake and sleep states. In summary, FM and HC show similar results when it comes to $\alpha_{0} - \alpha_{q+}$, $\alpha_{q-} - \alpha_{0}$, $\alpha_{q-} - \alpha_{q+}$, $D_{0}$, $D_{1}$, and $D_{2}$ measures, i.e., both groups exhibit significant differences between awake and sleep segments. Furthermore, neither of these groups the width of the spectrum $\alpha_{q-} - \alpha_{q+}$ but show distinct features when it comes to the right side of the parabola. In this measure, only healthy controls show differences between awake and sleep states.\\
The right side of the parabola ($\alpha_{q-} - \alpha_{0}$) is associated with heterogeneity in small singularities, and, in this work, with heterogeneity in smooth movements.  The results might suggest that the awake state presents more heterogeneity in smooth movements than sleep in healthy group. The lack of difference between these two states, for FM, could be associated with sleep dysfunctions, a common complaint in these patients \cite{Wolfe1990} and might be connected to this commonly known issue in the sleep of this patients, previously reported in literature \cite{Wolfe1990}.\\

\section{Conclusions}
Actigraphy data exhibited a multifractal pattern and it was possible to (indirectly) find differences between healthy and fibromyalgia group by multifractal indices. Multifractal techniques, when applied to actigraphy signal, were able to reveal patterns in movement activation and smoothness heterogeneity. The characterisation of such patterns represents a new way to evaluate activity data, bringing key information about motor behaviour in subjects with chronic pain.\\
In the case of this study, the distinct features of awake and sleep states were able to show dissimilarities between a healthy group and patients with fibromyalgia, suggesting that sleep disturbances, reported literature, could be evidenced by movement patterns and heterogeneity.

\section*{Acknowledgements}

This work was supported by the Spanish Ministry of Science and Innovation and European Regional Development Funds (MICINN PSI2010-19372) and by the Funda\c{c}\~ao de Amparo \`a Pesquisa do Estado da Bahia (FAPESB) (PRONEX 0006/2009). LGSF was supported by the Brazilian funding agency Coordena\c{c}\~ao de Aperfei\c{c}oamento de Pessoal de N\'ivel Superior (Capes). LGSF and JGVM were supported by grants from Brazilian National Council of Research and Development (CNPq) (206907/2014-1 and 201499/2012-6, respectively).



\bibliographystyle{elsarticle-num}

\bibliography{ref}


\end{document}